\documentclass[preprintnumbers]{revtex4}
\usepackage{amsfonts}
\usepackage{amsmath}
\usepackage{hyperref}
\usepackage{amssymb}
\usepackage[english]{babel}
\usepackage{graphicx}
\usepackage{epsfig}
\usepackage{bm}
\usepackage{verbatim}
\usepackage[utf8]{inputenc}
\usepackage{color}
\usepackage{xcolor}
\usepackage{subfig}
\usepackage[normalem]{ulem}
\usepackage{dsfont}
\usepackage{multirow}
\newcommand{\mathsym}[1]{{}}

\input{tcilatex}
\begin{document}

\title{$\Delta(27)$ framework for cobimaximal neutrino mixing models}
\author{A. E. C\'arcamo Hern\'andez}
\email{antonio.carcamo@usm.cl}
\affiliation{{\small Universidad T\'{e}cnica Federico Santa Mar\'{\i}a and Centro Cient%
\'{\i}fico-Tecnol\'{o}gico de Valpara\'{\i}so}\\
Casilla 110-V, Valpara\'{\i}so, Chile}
\author{I. de Medeiros Varzielas}
\email{ivo.de@udo.edu}
\affiliation{CFTP, Departamento de F\'{\i}sica, Instituto Superior T\'{e}cnico,
Universidade de Lisboa, Avenida Rovisco Pais 1, 1049 Lisboa, Portugal}
\date{\today }

\begin{abstract}
We propose a simple framework based on $\Delta(27)$ that leads to the successful
cobimaximal lepton mixing ansatz, thus providing a predictive explanation for leptonic mixing observables. We explore first the effective neutrino mass operators, then present a specific model realization based on type I seesaw, and also propose a model
with radiative 1-loop seesaw which features viable dark matter candidates.
\end{abstract}

\maketitle

\section{Introduction}

The experimental observation of 3 generations of fermions, with the
associated proliferation of parameters (the different masses and mixing
angles) constitute the flavour problem, in the Standard Model (SM) these
parameters remain free and are simply fitted to observations. Beyond SM
theories can be used to attempt explanations of the flavour problem, and be
used to predict (or postdict) these parameters, e.g. by providing relations
between mixing angles and the Dirac CP phase.

Neutrino oscillation experiments have now measured the leptonic mixing
angles with good precision and global fits provide an indication of the Dirac CP phase. The
currently observed leptonic mixing pattern can be successfully described by
the cobimaximal mixing pattern, which has recently received more attention 
\cite%
{Fukuura:1999ze,Miura:2000sx,Ma:2002ce,Ma:2015fpa,Ma:2016nkf,Damanik:2017jar,Ma:2017moj,Ma:2017trv,Grimus:2017itg,CarcamoHernandez:2017owh,CarcamoHernandez:2018hst,Ma:2019iwj,Ma:2019byo}%
.

Several extensions of the SM model with extended particle spectrum and
discrete flavour groups have been proposed to explain the pattern on lepton
masses and mixings. Among these, the discrete group $\Delta(27)$ has several
nice properties that make it interesting as a family symmetry - the
irreducible representations are a triplet and anti-triplet, plus 9 distinct
singlets. $\Delta(27)$ has been used widely in the literature, often in
association with CP symmetries \cite%
{Branco:1983tn,deMedeirosVarzielas:2006fc, Ma:2006ip,
Ma:2007wu,Bazzocchi:2009qg, deMedeirosVarzielas:2011zw, Varzielas:2012nn,Bhattacharyya:2012pi,Ferreira:2012ri,Ma:2013xqa,Nishi:2013jqa,Varzielas:2013sla,Aranda:2013gga,Varzielas:2013eta,Harrison:2014jqa,Ma:2014eka,Abbas:2014ewa,Abbas:2015zna,Varzielas:2015aua,Bjorkeroth:2015uou,Chen:2015jta,Vien:2016tmh,Hernandez:2016eod, Bjorkeroth:2016lzs, CarcamoHernandez:2017owh,deMedeirosVarzielas:2017sdv, Bernal:2017xat,CarcamoHernandez:2018iel,deMedeirosVarzielas:2018vab,CarcamoHernandez:2018hst,CarcamoHernandez:2018djj,Ma:2019iwj,Bjorkeroth:2019csz}%
.

The cobimaximal pattern for leptonic mixing is good explanation for the observed neutrino oscillation data. This pattern corresponds to a neutrino mass matrix of the form:
\begin{equation}
\widetilde{M}_{\nu}=\left( 
\begin{array}{ccc}
A & B & B^{*} \\ 
B & C & D \\ 
B^{*} & D & C^{*}
\end{array}%
\right),
\label{X}
\end{equation}
in the basis where the SM charged lepton mass matrix is diagonal. This pattern predicts $\theta_{13}\neq 0$, $\theta_{23}=\frac{\pi}{4}$ and $\delta_{CP}=-\frac{\pi}{2}$, which agrees well with the experimental data on neutrino oscillations. The pattern is called cobimaximal because it predicts the maximal allowed leptonic mixing in the 23 plane as well as maximal leptonic Dirac CP violating phase. It also corresponds to a generalized $\mu-\tau$ symmetry ~\cite{babu:2002dz,grimus:2003yn,King:2014nza}: 
\begin{equation}
P^T\widetilde{M}_{\nu}P=\left(\widetilde{M}_{\nu}\right)^{*} 
\end{equation} 
with 
\begin{equation}
P=\left( 
\begin{array}{ccc}
1 & 0 & 0 \\ 
0 & 0 & 1 \\ 
0 & 1 & 0
\end{array}%
\right).
\label{X}
\end{equation}
To obtain cobimaximal mixing, $\Delta(27)$ has been used recently in \cite{Ma:2019iwj}, and in \cite{CarcamoHernandez:2017owh,CarcamoHernandez:2018hst}.
We note however that in \cite{Ma:2019iwj}, soft breaking of $\Delta(27)$
is invoked, whereas in \cite%
{CarcamoHernandez:2017owh,CarcamoHernandez:2018hst} the breaking of $%
\Delta(27)$ employs a generic direction. In contrast, in this paper we will
show how $\Delta(27)$ is a good family symmetry to construct cobimaximal
models with the breaking of $\Delta(27)$ following natural directions that
are easy to obtain with the group. The layout of the paper is as follows. In
section \ref{models} we describe two $\Delta(27)$ flavour models that lead to
the cobimaximal mixing pattern. The implications of those models in lepton
masses and mixings are analysed in Section \ref{leptonmixins}. Our
conclusions are given in Section \ref{conclusions}. Appendix \ref{delta27}
contains a brief description of the $\Delta(27)$ discrete group.

\section{Models}

\label{models} At the effective level we intend to obtain a framework where,
in the model building basis for the fermions, the charged lepton Yukawa
matrix is diagonal together with a neutrino mass matrix which is
diagonalized by the cobimaximal ansatz. Considering the effective operators,
such a neutrino mass matrix can be achieved with the following Lagrangian: 
\begin{eqnarray}
\tciLaplace _{Y}^{\left( W\nu \right) } &=&\frac{\kappa _{1}}{\Lambda ^{3}}(%
\overline{l}_{L}h_{u}\phi _{23})(l_{L}^{C}h_{u}\phi _{23})+\frac{\kappa _{2}%
}{\Lambda ^{3}}(\overline{l}_{L}h_{u}\phi _{1})(l_{L}^{C}h_{u}\phi _{1})+%
\frac{\kappa _{3}}{\Lambda ^{3}}(\overline{l}_{L}h_{u}\phi
_{123})(l_{L}^{C}h_{u}\phi _{123})  \notag \\
&&+\frac{\kappa _{4}}{\Lambda ^{3}}\left[ (\overline{l}_{L}h_{u}\phi
_{1})(l_{L}^{C}h_{u}\phi _{123})+(\overline{l}_{L}h_{u}\phi
_{123})(l_{L}^{C}h_{u}\phi _{1})\right] +h.c.,
\end{eqnarray}%
which requires $\phi _{23}$ to be distinguished by e.g. a $Z_{2}$ from the
other two flavons, $\phi _{1}$ and $\phi _{123}$. This leads to a
cobimaximal form for the neutrino mass matrix, provided a CP symmetry is
imposed forcing the coefficients to be real, and that the following VEV
patterns for the $\Delta(27)$ triplets SM singlet scalar fields is considered: 
\begin{equation}
\left\langle \rho \right\rangle =v_{\rho }\left( 1,0,0\right) ,\hspace{1cm}%
\left\langle \phi _{1}\right\rangle =v_{1}\left( 1,0,0\right) ,\hspace{1cm}%
\left\langle \phi _{123}\right\rangle =v_{123}\left( 1,\omega ,\omega
^{2}\right) ,\hspace{1cm}\left\langle \phi _{23}\right\rangle =v_{23}\left(
0,1,-1\right) ,  \label{D27VEVS}
\end{equation}%
where $\omega =e^{i\frac{2\pi }{3}}$, and $\rho $ will be responsible for
ensuring the charged lepton Yukawa matrix is diagonal in this basis.
Cobimaximal mixing is motivating these VEVs, and in turn these special
directions are the motivation for realizing the models in a SUSY framework
with a $\Delta (27)$ family symmetry, as they can be easily obtained in
SUSY $\Delta (27)$ flavour models through F-term alignment mechanism \cite%
{Varzielas:2015aua} or D-term alignment mechanism \cite%
{deMedeirosVarzielas:2006fc}. This is in contrast with the somewhat generic $%
(r,e^{i\psi },e^{-i\psi })$ VEV employed in \cite%
{CarcamoHernandez:2017owh,CarcamoHernandez:2018hst}. We therefore consider
implicitly extensions of the minimal supersymmetric SM (MSSM), although for
our purposes, it is enough to assume that these VEV directions are obtained
e.g. through F-term alignment \cite{Varzielas:2015aua}, and that the Yukawa
Lagrangian arises from an holomorphic superpotential.

In the following subsections we are going to describe two specific models
where the cobimaximal mixing pattern is obtained, by adding 3 right-handed (RH)
neutrinos.

\subsection{Model 1.}

In this supersymmetric model, the full symmetry $\mathcal{G}$ experiences a
two-step spontaneous breaking: 
\begin{eqnarray}
&&\mathcal{G}=SU(3)_{C}\times SU(2)_{L}\times U\left( 1\right) _{Y}\times
\Delta(27) \times Z_{2}\times Z_{10}  \notag \\
&&\hspace{35mm}\Downarrow \Lambda _{int}  \notag \\[3mm]
&&\hspace{15mm}SU\left( 3\right) _{C}\times SU\left( 2\right) _{L}\times
U\left( 1\right) _{Y}  \notag \\[3mm]
&&\hspace{35mm}\Downarrow v  \notag \\[3mm]
&&\hspace{23mm}SU\left( 3\right) _{C}\otimes U\left( 1\right) _{em}  \label{Group}
\end{eqnarray}
%where $\Lambda _{int}$ is the scale of breaking of the discrete groups, whereas $v=246$ GeV is the electroweak symmetry breaking scale. 
It is assumed that the discrete groups are spontaneously broken at an energy
scale $\Lambda _{int}$ much larger than the electroweak symmetry breaking scale $v=246$ GeV.
In the supersymmetric model under consideration, the scalar sector of
MSSM is extended by the inclusion of several gauge singlet scalar fields,
whereas the fermion sector is enlarged by considering three very heavy
RH Majorana neutrinos. Such heavy RH Majorana neutrinos
are crucial for mediating a type I seesaw mechanism that produces the tiny
values of the light active neutrino masses. The inclusion of the gauge
singlet scalars is necessary for the implementation of the
Froggat-Nielsen mechanism that produces the SM charged lepton mass hierarchy
and allows to build the neutrino Yukawa terms invariant under the symmetries
of the model, that give rise to a predictive cobimaximal neutrino mass
matrix texture. The $\Delta(27) \times Z_{2}\times Z_{10}$ assignments of
fermions and scalars in our model are shown in Table \ref{Themodel}. 
\begin{table}[th]
\centering 
\begin{tabular}{|c|c|c|c|c|c|c|c|c|c|c|c|c|c|c|}
\hline\hline
& $l_{L}$ & $l_{1R}$ & $l_{2R}$ & $l_{3R}$ & $N_{1R}$ & $N_{2R}$ & $N_{3R}$
& $h_{u}$ & $h_{d}$ & $\sigma$ & $\rho$ & $\phi _{1}$ & $\phi _{123}$ & $%
\phi _{23}$ \\ \hline
$\Delta(27)$ & $\mathbf{3}$ & $\mathbf{1}_{\mathbf{0,0}}$ & $\mathbf{1 }_{%
\mathbf{0,1}}$ & $\mathbf{1}_{\mathbf{0,2}}$ & $\mathbf{1}_{\mathbf{0,0}}$ & 
$\mathbf{1}_{\mathbf{0,0}}$ & $\mathbf{1}_{\mathbf{0,0}}$ & $\mathbf{1}_{%
\mathbf{\ 0,0}}$ & $\mathbf{1}_{\mathbf{0,0}}$ & $\mathbf{1}_{\mathbf{0,0}}$
& $\mathbf{3}$ & $\mathbf{3}$ & $\mathbf{3}$ & $\mathbf{3}$ \\ \hline
$Z_{2}$ & $0$ & $1$ & $1$ & $1$ & $0$ & $0$ & $0$ & $0$ & $0$ & $0$ & $1$ & $%
0$ & $0$ & $0$ \\ \hline
$Z_{10}$ & $0$ & $5$ & $4$ & $2$ & $0$ & $0$ & $5$ & $0$ & $0$ & $-1$ & $0$
& $0$ & $0$ & $5$ \\ \hline
\end{tabular}%
\caption{Leptonic and scalar field assignments under the $\Delta(27)\times
Z_2\times Z_{10}$ symmetry.}
\label{Themodel}
\end{table}\newline
Notice that in Table \ref{Themodel}, the numbers in boldface correspond to the $\Delta(27)$ representations
and the $Z_{N}$ charges are written in additive notation.

In this model, the $\Delta(27)$ discrete flavor symmetry is necessary to get
a predictive lepton sector through the special VEV directions in Eq. (\ref%
{D27VEVS}). The $Z_{2}$ symmetry separates the $\Delta(27)$ scalar triplet $%
\rho $ that participates in the charged lepton Yukawa interactions from the
ones ($\phi _{1}$, $\phi _{123}$, $\phi _{23}$) appearing in the neutrino
Yukawa terms, thus allowing to treat these sectors independently. The $%
Z_{10} $ implements the Froggat-Nielsen mechanism that produces the SM
charged lepton mass hierarchy, and also distinguishes the $\Delta(27)$
scalar triplet $\phi _{23}$ from the $\Delta \left( 27\right) $ scalar
triplets $\phi _{1}$ and $\phi _{123}$.

%thus giving rise to a predictive neutrino mass matrix consistent with the cobimaximal mixing pattern.

Since the spontaneous breaking of the $Z_{10}$ symmetry produces the SM
charged fermion mass hierarchy, we set the vacuum expectation values (VEVs)
of the different gauge singlet scalars as follows:

\begin{equation}
v_{\rho }\sim v_{\sigma }\sim v_{1}\sim v_{123}\sim v_{23}\sim \lambda
\Lambda ,  \label{VEVhierarchy}
\end{equation}

where $\lambda =\sin \theta _{13}$, being $\theta _{13}$ is the reactor
mixing angle and $\Lambda $ the model cutoff, which can be interpreted as
the scale of the UV completion of the model, e.g. the masses of the
Froggatt-Nielsen messenger fields.

The Yukawa terms for the lepton sector invariant under the aforementioned
symmetries are:

\begin{equation}
\tciLaplace _{Y}^{\left( l\right) }=y_{1}^{\left( l\right) }\left( \overline{%
l}_{L}\rho h_{d}\right) _{\mathbf{1}_{\mathbf{0,0}}}l_{1R}\frac{\sigma ^{8}}{%
\Lambda ^{9}}+y_{2}^{\left( l\right) }\left( \overline{l}_{L}\rho
h_{d}\right) _{\mathbf{1}_{\mathbf{0,2}}}l_{2R}\frac{\sigma ^{4}}{\Lambda
^{5}}+y_{3}^{\left( l\right) }\left( \overline{l}_{L}\rho h_{d}\right) _{%
\mathbf{1}_{\mathbf{0,1}}}l_{3R}\frac{\sigma ^{2}}{\Lambda ^{3}}+h.c.,
\label{Lyl}
\end{equation}
\begin{eqnarray}
\tciLaplace _{Y}^{\left( \nu \right) } &=&y_{1}^{\left( \nu \right) }\left( 
\overline{l}_{L}\phi _{1}h_{u}\right) _{\mathbf{1}_{\mathbf{0,0}}}N_{1R}%
\frac{1}{\Lambda }+y_{2}^{\left( \nu \right) }\left( \overline{l}_{L}\phi
_{123}h_{u}\right) _{\mathbf{1}_{\mathbf{0,0}}}N_{2R}\frac{1}{\Lambda } 
\notag \\
&&+y_{3}^{\left( \nu \right) }\left( \overline{l}_{L}\phi _{123}h_{u}\right)
_{\mathbf{1}_{\mathbf{0,0}}}N_{1R}\frac{1}{\Lambda }+y_{4}^{\left( \nu
\right) }\left( \overline{l}_{L}\phi _{1}h_{u}\right) _{\mathbf{1}_{\mathbf{%
0,0}}}N_{2R}\frac{1}{\Lambda }+y_{2}^{\left( \nu \right) }\left( \overline{l}%
_{L}\phi _{23}h_{u}\right) _{\mathbf{1}_{\mathbf{0,0}}}N_{3R}\frac{1}{%
\Lambda }  \notag \\
&&+m_{N_{1}}\overline{N}_{1R}N_{1R}^{C}+m_{N_{2}}\overline{N}%
_{2R}N_{2R}^{C}+m_{N_{3}}\overline{N}_{3R}N_{3R}^{C}+m_{N_{4}}\left( 
\overline{N}_{1R}N_{2R}^{C}+\overline{N}_{2R}N_{1R}^{C}\right) +h.c..
\label{Lynu1}
\end{eqnarray}
This is the most general form. We can without loss of generality change the
RH neutrino basis such that we choose states where the RH
neutrinos $N_{1}$ and $N_{2}$ don't mix (RH neutrino diagonal mass basis) or
we can choose states where $N_{1}$ couples only to $\phi _{1}$ and $N_{2}$
couples only to $\phi _{123}$ (RH neutrino flavon basis). In any case, the
effective low energy neutrinos will have a term of the form $\frac{\kappa
_{4}}{\Lambda ^{3}}\left[ (\overline{l}_{L}h_{u}\phi
_{1})(l_{L}^{C}h_{u}\phi _{123})+(\overline{l}_{L}h_{u}\phi
_{123})(l_{L}^{C}h_{u}\phi _{1})\right] $, as required.

\subsection{Model 2.}

This model is very similar to model 1. The crucial difference here is that
the light active neutrino masses are generated from a one loop level
radiative seesaw mechanism instead of the tree level type I seesaw mechanism
of model 1. To implement such radiative seesaw mechanism in the model 2, we
add an extra preserved $Z_{2}^{\prime }$ symmetry, under which the right
handed Majorana neutrinos and an extra inert scalar singlet $\varphi $,
transforming as a $\Delta(27)$ trivial singlet, will be $Z_{2}^{\prime }$
charged, as follows:
\begin{equation}
\left(\varphi,N_{iR}\right)\rightarrow -\left(\varphi,N_{iR}\right),\hspace{1cm}i=1,2,3.
\end{equation}
The whole discrete group of this model will be $\Delta(27) \times
Z_{2}\times Z_{10}\times Z_{2}^{\prime }$, where the $\Delta(27) \times
Z_{2}\times Z_{10}$ group is spontaneously broken as in model 1 whereas the $%
Z_{2}^{\prime }$ symmetry is preserved. Due to the preserved $Z_{2}^{\prime
} $ symmetry, our model has scalar and fermionic dark matter candidates. The
scalar dark matter candidates will be the lightest of $\func{Re}\left(
\varphi \right) $ and $\func{Im}\left( \varphi \right) $, while the
fermionic dark matter candidate will be the lightest of the RH
Majorana neutrinos. The resulting implications of such model in Dark model
will be the same as in the model of Ref. \cite{Bernal:2017xat}, which makes
our model consistent with dark matter constraints.

The SM charged lepton Yukawa terms will be the same as in model 1, whereas
the neutrino Yukawa interactions take the form:

\begin{eqnarray}
\tciLaplace _{Y}^{\left( \nu \right) } &=&y_{1}^{\left( \nu \right) }\left( 
\overline{l}_{L}\phi _{1}h_{u}\right) _{\mathbf{1}_{\mathbf{0,0}}}N_{1R}%
\frac{\varphi }{\Lambda ^{2}}+y_{2}^{\left( \nu \right) }\left( \overline{l}%
_{L}\phi _{123}h_{u}\right) _{\mathbf{1}_{\mathbf{0,0}}}N_{2R}\frac{\varphi 
}{\Lambda ^{2}}  \notag \\
&&+y_{3}^{\left( \nu \right) }\left( \overline{l}_{L}\phi _{123}h_{u}\right)
_{\mathbf{1}_{\mathbf{0,0}}}N_{1R}\frac{\varphi }{\Lambda ^{2}}%
+y_{4}^{\left( \nu \right) }\left( \overline{l}_{L}\phi _{1}h_{u}\right) _{%
\mathbf{1}_{\mathbf{0,0}}}N_{2R}\frac{\varphi }{\Lambda ^{2}}+y_{2}^{\left(
\nu \right) }\left( \overline{l}_{L}\phi _{23}h_{u}\right) _{\mathbf{1}_{%
\mathbf{0,0}}}N_{3R}\frac{\varphi }{\Lambda ^{2}}  \notag \\
&&+m_{N_{1}}\overline{N}_{1R}N_{1R}^{C}+m_{N_{2}}\overline{N}%
_{2R}N_{2R}^{C}+m_{N_{3}}\overline{N}_{3R}N_{3R}^{C}+m_{N_{4}}\left( 
\overline{N}_{1R}N_{2R}^{C}+\overline{N}_{2R}N_{1R}^{C}\right) +h.c..
\label{Lynu2}
\end{eqnarray}

\section{Lepton masses and mixings}

\label{leptonmixins}

\subsection{Model 1.}

After the discrete groups are spontaneously broken, we get the following
charged lepton and neutrino Yukawa terms: 
\begin{equation}
\tciLaplace _{Y}^{\left( l\right) }=y_{1}^{\left( l\right) }\overline{l}%
_{1L}h_{d}l_{1R}\frac{v_{\rho }v_{\sigma }^{8}}{\Lambda ^{9}}+y_{2}^{\left(
l\right) }\overline{l}_{L}h_{d}l_{2R}\frac{v_{\rho }v_{\sigma }^{3}}{\Lambda
^{4}}+y_{3}^{\left( l\right) }\overline{l}_{L}h_{d}l_{3R}\frac{v_{\rho
}v_{\sigma }^{2}}{\Lambda ^{3}}+h.c.,
\end{equation}%
\begin{eqnarray}
\tciLaplace _{Y}^{\left( \nu \right) } &=&y_{1}^{\left( \nu \right) }%
\overline{l}_{1L}h_{u}N_{1R}\frac{v_{1}}{\Lambda }+y_{2}^{\left( \nu \right)
}\left( \overline{l}_{1L}+\omega \overline{l}_{2L}+\omega ^{2}\overline{l}%
_{3L}\right) h_{u}N_{2R}\frac{v_{123}}{\Lambda }  \notag \\
&&+y_{3}^{\left( \nu \right) }\left( \overline{l}_{1L}+\omega \overline{l}%
_{2L}+\omega ^{2}\overline{l}_{3L}\right) h_{u}N_{1R}\frac{v_{123}}{\Lambda }%
+y_{4}^{\left( \nu \right) }h_{u}N_{2R}\frac{v_{1}}{\Lambda }+y_{2}^{\left(
\nu \right) }\left( \overline{l}_{2L}-\overline{l}_{3L}\right) h_{u}N_{3R}%
\frac{v_{23}}{\Lambda }  \notag \\
&&+m_{N_{1}}\overline{N}_{1R}N_{1R}^{C}+m_{N_{2}}\overline{N}%
_{2R}N_{2R}^{C}+m_{N_{3}}\overline{N}_{3R}N_{3R}^{C}+m_{N_{4}}\left( 
\overline{N}_{1R}N_{2R}^{C}+\overline{N}_{2R}N_{1R}^{C}\right) +h.c..
\end{eqnarray}%
Consequently, the SM charged lepton mass matrix is diagonal with the charged
lepton masses given by: 
\begin{equation}
m_{e}=y_{1}^{\left( l\right) }\frac{v_{\rho }v_{\sigma }^{8}v_{h_{d}}}{\sqrt{%
2}\Lambda ^{9}}=a_{1}^{\left( l\right) }\lambda ^{9}\frac{v}{\sqrt{2}},%
\hspace{1cm}m_{\mu }=y_{2}^{\left( l\right) }\frac{v_{\rho }v_{\sigma
}^{4}v_{h_{d}}}{\sqrt{2}\Lambda ^{5}}=a_{2}^{\left( l\right) }\lambda ^{5}%
\frac{v}{\sqrt{2}},\hspace{1cm}m_{\tau }=y_{3}^{\left( l\right) }\frac{%
v_{\rho }v_{\sigma }^{2}v_{h_{d}}}{\sqrt{2}\Lambda ^{3}}=a_{3}^{\left(
l\right) }\lambda ^{3}\frac{v}{\sqrt{2}}
\end{equation}%
where $a_{1}^{\left( l\right) }$, $a_{2}^{\left( l\right) }$ and $%
a_{3}^{\left( l\right) }$ are real $\mathcal{O}(1)$ dimensionless parameters
and we have assumed that $v_{h_{d}}\sim v/\sqrt{2}$, being $v=246$ GeV the
electroweak symmetry breaking scale.

In what regards the neutrino sector, we find that the full $6\times 6$
neutrino mass matrix read: 
\begin{equation}
M_{\nu }=\left( 
\begin{array}{cc}
0_{3\times 3} & M_{\nu D} \\ 
M_{\nu D}^{T} & M_{R}%
\end{array}%
\right),
\end{equation}%
where the Dirac and Majorana neutrino mass matrices are given by: 
\begin{equation}
M_{\nu D}=\left( 
\begin{array}{ccc}
A_{1}+A_{2} & B_{1}+B_{2} & 0 \\ 
\omega A_{2} & \omega B_{1} & C \\ 
\omega ^{2}A_{2} & \omega ^{2}B_{1} & -C%
\end{array}%
\right) ,\hspace{1.5cm}M_{R}=\left( 
\begin{array}{ccc}
m_{N_{1}} & m_{N_{4}} & 0 \\ 
m_{N_{4}} & m_{N_{2}} & 0 \\ 
0 & 0 & m_{N_{3}}%
\end{array}%
\right),
\end{equation}%
with: 
\begin{eqnarray}
A_{1} &=&y_{1}^{\left( \nu \right) }\frac{v_{1}}{\Lambda }\frac{v_{h_{u}}}{%
\sqrt{2}},\hspace{1.5cm}A_{2}=y_{2}^{\left( \nu \right) }\frac{v_{123}}{%
\Lambda }\frac{v_{h_{u}}}{\sqrt{2}},  \notag \\
B_{1} &=&y_{2}^{\left( \nu \right) }\frac{v_{123}}{\Lambda }\frac{v_{h_{u}}}{%
\sqrt{2}},\hspace{1.5cm}B_{2}=y_{4}^{\left( \nu \right) }\frac{v_{1}}{%
\Lambda }\frac{v_{h_{u}}}{\sqrt{2}},\hspace{1.5cm}C=y_{2}^{\left( \nu
\right) }\frac{v_{23}}{\Lambda }\frac{v_{h_{u}}}{\sqrt{2}}.
\end{eqnarray}
Assuming that the Majorana neutrino masses are much larger than the
electroweak symmetry breaking scale, the light active neutrino masses will
be generated from a type I seesaw mechanism. Thus, the light active neutrino
mass matrix takes the form: 
\begin{equation}
\widetilde{M}_{\nu }=M_{\nu D}M_{R}^{-1}M_{\nu D}^{T}=\left( 
\begin{array}{ccc}
a & d\omega & d\omega ^{2} \\ 
d\omega & be^{i\theta } & c \\ 
d\omega ^{2} & c & be^{-i\theta }%
\end{array}%
\right),\label{Mnu}
\end{equation}
where: 
\begin{eqnarray}
a &=&X\left( A_{1}+A_{2}\right) ^{2}+2W\left( A_{1}+A_{2}\right) \left(
B_{1}+B_{2}\right) +Y\left( B_{1}+B_{2}\right) ^{2},  \notag \\
d &=&\left( XA_{2}+WB_{1}\right) \left( A_{1}+A_{2}\right) +\left(
YB_{1}+WA_{2}\right) \left( B_{1}+B_{2}\right) ,  \notag \\
c &=&-ZC^{2}+XA_{2}^{2}+2WA_{2}B_{1}+YB_{1}^{2},  \notag \\
b &=&\left\vert ZC^{2}+\omega ^{2}\left(
XA_{2}^{2}+2WA_{2}B_{1}+YB_{1}^{2}\right) \right\vert ,  \notag \\
\theta &=&\arg \left( ZC^{2}+\omega ^{2}\left(
XA_{2}^{2}+2WA_{2}B_{1}+YB_{1}^{2}\right) \right),
\end{eqnarray}
\begin{equation}
X=\frac{m_{N_{2}}}{m_{N_{1}}m_{N_{2}}-m_{N_{4}}^{2}},\hspace{1.5cm}Y=\frac{%
m_{N_{1}}}{m_{N_{1}}m_{N_{2}}-m_{N_{4}}^{2}},\hspace{1.5cm}W=-\frac{m_{N_{4}}%
}{m_{N_{1}}m_{N_{2}}-m_{N_{4}}^{2}},\hspace{1.5cm}Z=\frac{1}{m_{N_{3}}}.
\end{equation}
Such light active neutrino mass matrix features a cobimaximal mixing pattern
and is diagonalized by the rotation matrix: 
\begin{equation}
U=\left( 
\begin{array}{ccc}
\cos \alpha _{12}\cos \alpha _{13} & -\cos \alpha _{13}\sin \alpha _{12} & 
-\sin \alpha _{13} \\ 
\frac{\sin \alpha _{12}-i\cos \alpha _{12}\sin \alpha _{13}}{\sqrt{2}} & 
\frac{\cos \alpha _{12}+i\sin \alpha _{12}\sin \alpha _{13}}{\sqrt{2}} & -%
\frac{i\cos \alpha _{13}}{\sqrt{2}} \\ 
\frac{\sin \alpha _{12}+i\cos \alpha _{12}\sin \alpha _{13}}{\sqrt{2}} & 
\frac{\cos \alpha _{12}-i\sin \alpha _{12}\sin \alpha _{13}}{\sqrt{2}} & 
\frac{i\cos \alpha _{13}}{\sqrt{2}}%
\end{array}%
\right),\label{Ucobimaximal}
\end{equation}
as follows: 
\begin{equation}
U^{T}\widetilde{M}_{\nu }U=\left( 
\begin{array}{ccc}
m_{1} & 0 & 0 \\ 
0 & m_{2} & 0 \\ 
0 & 0 & m_{3}%
\end{array}%
\right).
\label{Mnudiag}
\end{equation}
From Eq. (\ref{Mnudiag}), we find that the parameters $a$, $b$, $c$, $d$ and 
$\theta $ of the light active neutrino mass matrix $\widetilde{M}_{\nu }$
are given by the following relations: 
\begin{table}[t]
{\footnotesize \ }
\par
\begin{center}
{\footnotesize \ \renewcommand{\arraystretch}{1} 
\begin{tabular}{c|c||c|c|c|c}
\hline
\multirow{2}{*}{Observable} & \multirow{2}{*}{\parbox{7em}{Model\\ value} }
& \multicolumn{4}{|c}{Neutrino oscillation global fit values (NH)} \\ 
\cline{3-6}
&  & Best fit $\pm 1\sigma$ \cite{deSalas:2017kay} & Best fit $\pm 1\sigma$ 
\cite{Esteban:2018azc} & $3\sigma$ range \cite{deSalas:2017kay} & $3\sigma$
range \cite{Esteban:2018azc} \\ \hline\hline
$\Delta m_{21}^{2}$ [$10^{-5}$eV$^{2}$] & $7.40$ & $7.55_{-0.16}^{+0.20}$ & $%
7.39_{-0.20}^{+0.21}$ & $7.05-8.14$ & $6.79-8.01$ \\ \hline
$\Delta m_{31}^{2}$ [$10^{-3}$eV$^{2}$] & $2.49$ & $2.50\pm 0.03$ & $%
2.525_{-0.032}^{+0.033}$ & $2.41-2.60$ & $2.427-2.625$ \\ \hline
$\theta _{12}(^{\circ })$ & $34.5$ & $34.5_{-1.0}^{+1.2}$ & $%
33.82_{-0.76}^{+0.78}$ & $31.5-38.0$ & $31.61-36.27$ \\ \hline
$\theta _{13}(^{\circ })$ & $8.45$ & $8.45_{-0.14}^{+0.16}$ & $8.61\pm 0.13$
& $8.0-8.9$ & $8.22-8.99$ \\ \hline
$\theta _{23}(^{\circ })$ & $45$ & $47.7_{-1.7}^{+1.2}$ & $%
49.6_{-1.2}^{+1.0} $ & $41.8-50.7$ & $40.3-52.4$ \\ \hline
$\delta _{CP}(^{\circ })$ & $-90$ & $218_{-27}^{+38}$ & $215_{-29}^{+40}$ & $%
157-349$ & $125-392$ \\ \hline\hline
\end{tabular}
{\normalsize \ }}
\end{center}
\par
{\footnotesize \ }
\caption{Model and experimental values of
the neutrino mass squared splittings, leptonic mixing angles, and $CP$%
-violating phase. The experimental values are taken from Refs.~\protect\cite%
{deSalas:2017kay,Esteban:2018azc}.}
\label{tab:neutrinos-NH}
\end{table}
\begin{eqnarray}
a &=&m_{3}\sin ^{2}\alpha _{13}+\cos ^{2}\alpha _{13}\left( m_{2}\sin
^{2}\alpha _{12}+m_{1}\cos ^{2}\alpha _{12}\right) ,  \notag \\
d &=&\frac{\cos \alpha _{13}\left( \left( m_{1}-m_{2}\right) \sin 2\alpha
_{12}-2i\sin \alpha _{13}\left( m_{2}\sin ^{2}\alpha _{12}+m_{1}\cos
^{2}\alpha _{12}-m_{3}\right) \right) }{2\sqrt{2}},  \notag \\
c &=&\frac{1}{8}\left( 4m_{3}\cos ^{2}\alpha _{13}-m_{1}\left( 2\cos 2\alpha
_{13}\cos ^{2}\alpha _{12}+\cos 2\alpha _{12}-3\right) +m_{2}\left( \cos
2\alpha _{12}-2\sin ^{2}\alpha _{12}\cos 2\alpha _{13}+3\right) \right) 
\notag \\
b &=&\left\vert \frac{1}{8}\left( -4m_{3}\cos ^{2}\alpha _{13}+4m_{1}\left(
\sin \alpha _{12}-i\sin \alpha _{13}\cos \alpha _{12}\right)
{}^{2}+4m_{2}\left( \cos \alpha _{12}+i\sin \alpha _{12}\sin \alpha
_{13}\right) {}^{2}\right) \right\vert ,  \notag \\
\theta &=&\arg \left[ \frac{1}{8}\left( -4m_{3}\cos ^{2}\alpha
_{13}+4m_{1}\left( \sin \alpha _{12}-i\sin \alpha _{13}\cos \alpha
_{12}\right) {}^{2}+4m_{2}\left( \cos \alpha _{12}+i\sin \alpha _{12}\sin
\alpha _{13}\right) {}^{2}\right) \right].
\end{eqnarray}
The leptonic mixing angles and the CP phase arising from the cobimaximal
light active neutrino mass matrix $\widetilde{M}_{\nu }$ read: 
\begin{equation}
\theta _{13}=\alpha _{13},\hspace{1.5cm}\theta _{12}=\alpha _{12},\hspace{%
1.5cm}\theta _{23}=\frac{\pi }{4},\hspace{1.5cm}\delta _{C`P}=-\frac{\pi }{2}.
\end{equation}
The physical observables of the neutrino sector, i.e., the three leptonic
mixing angles, the CP phase and the neutrino mass squared splittings for the normal mass hierarchy (NH) can be
very well reproduced, as shown in Table \ref{tab:neutrinos-NH}, starting
from the following benchmark point: 
\begin{equation}
a\simeq 3.53\mbox{meV},\hspace{1cm}b\simeq 21.51\mbox{meV},\hspace{1cm}c\simeq 27.50\mbox{meV},\hspace{1cm}%
d\simeq 5.69\mbox{meV},\hspace{1cm}%
\theta \simeq 178.39^{\circ }.\label{Parameterfit-NH}
\end{equation}
This shows that our predictive model successfully describes the current
neutrino oscillation experimental data. Notice that with only five effective
parameters, i.e., $a$, $b$, $c$, $d$ and $\theta$, we can successfully
reproduce the experimental values of the six physical observables of the
neutrino sector: the neutrino mass squared differences, the leptonic mixing
angles and the leptonic CP phase.

\subsection{Model 2}

The resulting light active neutrino mass matrix arising from radiative
seesaw mechanism takes the form:

\begin{equation}
\widetilde{M}_{\nu }=\left( 
\begin{array}{ccc}
a & d\omega & d\omega ^{2} \\ 
d\omega & be^{i\theta } & c \\ 
d\omega ^{2} & c & be^{-i\theta }%
\end{array}%
\right)
\end{equation}

where $a,b,c,d, \theta$ are given by different expressions in terms of the high energy parameters
\begin{eqnarray}
a &=&X\left( A_{1}+A_{2}\right) ^{2}+2W\left( A_{1}+A_{2}\right) \left(
B_{1}+B_{2}\right) +Y\left( B_{1}+B_{2}\right) ^{2},  \notag \\
d &=&\left( XA_{2}+WB_{1}\right) \left( A_{1}+A_{2}\right) +\left(
YB_{1}+WA_{2}\right) \left( B_{1}+B_{2}\right) ,  \notag \\
c &=&-ZC^{2}+XA_{2}^{2}+2WA_{2}B_{1}+YB_{1}^{2},  \notag \\
b &=&\left\vert ZC^{2}+\omega ^{2}\left(
XA_{2}^{2}+2WA_{2}B_{1}+YB_{1}^{2}\right) \right\vert ,  \notag \\
\theta &=&\arg \left( ZC^{2}+\omega ^{2}\left(
XA_{2}^{2}+2WA_{2}B_{1}+YB_{1}^{2}\right) \right)
\end{eqnarray}%
where the parameters $X$, $Y$, $W$ will be loop functions depending on the
masses of the $Z_{2}^{\prime }$ odd scalars $\func{Re}\left( \varphi \right) 
$ and $\func{Im}\left( \varphi \right) $ and $Z_{2}^{\prime }$ odd right
handed Majorana neutrinos $N_{1R}$, $N_{2R}$ and $N_{3R}$ similar as in 
\cite{Bernal:2017xat,Ma:2006km}. Given that the light active neutrino
mass matrix in model 2 is of the same form as in model 1, the resulting
predictions in low energy neutrino observables will be the same in these
models and the benchmark point above for $a,b,c,d, \theta$ reproduces well the observed values.

\section{Conclusions}

\label{conclusions}

We have proposed a framework where the cobimaximal mixing pattern is
successfully realized, based on the $\Delta(27)$ family symmetry broken by
specific directions which are easy to obtain with this group. In this
framework, due to the $\Delta(27)$ breaking pattern and in the
model-building basis, the Standard Model charged lepton mass matrix is
diagonal whereas the light active neutrino mass matrix features a
cobimaximal mixing pattern. This generates the experimental values of the
neutrino mass squared splittings, the leptonic mixing angles and the
leptonic Dirac CP violating phase.

We present two example models within this framework, where $\Delta(27)$ is
supplemented by other auxiliary cyclic symmetries. In the first model, the
small masses for the light active neutrinos are produced from a tree-level
type I seesaw mechanism mediated by three heavy right-handed Majorana
neutrinos. In the second model, the light active neutrino masses arise from
a radiative seesaw mechanism where three right-handed Majorana neutrinos and
a gauge singlet scalar are charged under a preserved $Z_{2}^{\prime }$
symmetry, thus having viable dark matter candidates in conjunction with the
viable cobimaximal mixing structure.

\section*{Acknowledgements}

IdMV acknowledges funding from Funda\c{c}\~{a}o para a Ci\^{e}ncia e a
Tecnologia (FCT) through the contract IF/00816/2015 and by Funda\c{c}\~{a}o
para a Ci\^{e}ncia e a Tecnologia (FCT, Portugal) through the project
CFTP-FCT Unit 777 (UID/FIS/00777/2013) which is partially funded through
POCTI (FEDER), COMPETE, QREN and EU. AECH has received funding from Fondecyt
(Chile), Grants No.~1170803, CONICYT PIA/Basal FB0821. A.E.C.H thanks the
Instituto Superior T\'{e}cnico, Universidade de Lisboa for hospitality,
where this work was done.

\appendix

\section{The $\Delta (27)$ discrete group}

\label{delta27} The $\Delta (27)$ discrete group has the following 11
irreducible representations: one triplet $\mathbf{3}$, one antitriplet $%
\overline{\mathbf{3}}$ and nine singlets $\mathbf{1}_{k,l}$ ($k,l=0,1,2$),
where $k$ and $l$ identify how the singlets transform under order 3 generators, corresponding to a $Z_{3}$ and $Z_{3}^{\prime }$ subgroups of $\Delta(27)$.
\begin{eqnarray}
\mathbf{3}\otimes \mathbf{3} &=&\overline{\mathbf{3}}_{S_{1}}\oplus 
\overline{\mathbf{3}}_{S_{2}}\oplus \overline{\mathbf{3}}_{A}  \notag \\
\overline{\mathbf{3}}\otimes \overline{\mathbf{3}} &=&\mathbf{3}%
_{S_{1}}\otimes \mathbf{3}_{S_{2}}\oplus \mathbf{3}_{A}  \notag \\
\mathbf{3}\otimes \overline{\mathbf{3}} &=&\sum_{r=0}^{2}\mathbf{1}%
_{r,0}\oplus \sum_{r=0}^{2}\mathbf{1}_{r,1}\oplus \sum_{r=0}^{2}\mathbf{1}%
_{r,2}  \notag \\
\mathbf{1}_{k,\ell }\otimes \mathbf{1}_{k^{\prime },\ell ^{\prime }} &=&%
\mathbf{1}_{k+k^{\prime }mod3,\ell +\ell ^{\prime }mod3} \\
&&
\end{eqnarray}%
Denoting $\left( x_{1},y_{1},z_{1}\right) $ and $\left(
x_{2},y_{2},z_{2}\right) $ as the basis vectors for two $\Delta (27)$%
-triplets $\mathbf{3}$, one finds: 
\begin{eqnarray}
\left( \mathbf{3}\otimes \mathbf{3}\right) _{\overline{\mathbf{3}}_{S_{1}}}
&=&\left( x_{1}y_{1},x_{2}y_{2},x_{3}y_{3}\right) ,  \notag
\label{triplet-vectors} \\
\left( \mathbf{3}\otimes \mathbf{3}\right) _{\overline{\mathbf{3}}_{S_{2}}}
&=&\frac{1}{2}\left(
x_{2}y_{3}+x_{3}y_{2},x_{3}y_{1}+x_{1}y_{3},x_{1}y_{2}+x_{2}y_{1}\right) , 
\notag \\
\left( \mathbf{3}\otimes \mathbf{3}\right) _{\overline{\mathbf{3}}_{A}} &=&%
\frac{1}{2}\left(
x_{2}y_{3}-x_{3}y_{2},x_{3}y_{1}-x_{1}y_{3},x_{1}y_{2}-x_{2}y_{1}\right) , 
\notag \\
\left( \mathbf{3}\otimes \overline{\mathbf{3}}\right) _{\mathbf{1}_{r,0}}
&=&x_{1}y_{1}+\omega ^{2r}x_{2}y_{2}+\omega ^{r}x_{3}y_{3},  \notag \\
\left( \mathbf{3}\otimes \overline{\mathbf{3}}\right) _{\mathbf{1}_{r,1}}
&=&x_{1}y_{2}+\omega ^{2r}x_{2}y_{3}+\omega ^{r}x_{3}y_{1},  \notag \\
\left( \mathbf{3}\otimes \overline{\mathbf{3}}\right) _{\mathbf{1}_{r,2}}
&=&x_{1}y_{3}+\omega ^{2r}x_{2}y_{1}+\omega ^{r}x_{3}y_{2},
\end{eqnarray}%
where $r=0,1,2$ and $\omega =e^{i\frac{2\pi }{3}}$.

\end{document}